# Wide-field Hyperspectral Optical Microscopy for Rapid Characterization of Two-Dimensional Semiconductors and Heterostructures

*Zhenghan Peng* [1], *Adeyemi Uthman* [1], *Zhepeng Zhang* [1], *Anh Tuan Hoang* [1], *Xiang Zhu* [1], *Eric Pop* [1,2,3], *Andrew J. Mannix* [1,4,*]

1. Department of Materials Science & Engineering, Stanford University, Stanford, CA 94305, USA
2. Department of Electrical Engineering, Stanford University, Stanford, CA 94305, USA
3. Department of Applied Physics, Stanford University, Stanford, CA 94305, USA
4. Stanford Institute for Materials and Energy Sciences, SLAC National Accelerator Laboratory, Menlo Park, CA 94025, USA

*Corresponding author: A.J.M., ajmannix@stanford.edu

**Abstract**

Electronic and optoelectronic applications of two-dimensional (2D) semiconductors demand precise control over material quality, including thickness, composition, doping, and defect density. Conventional benchmarking methods (e.g., charge transport, confocal mapping, electron or scanning probe microscopy) are slow, perturb sample quality, or involve trade-offs between speed, resolution, and scan area. To accelerate assessment of 2D semiconductors, we demonstrate a broadband, wide-field hyperspectral optical microscope for 2D materials (2D-HOM) that rapidly captures a spatial-spectral data cube within seconds. The data cube includes x-y spatial coordinate (a 300 × 300 $\mu m^2$ field, with ~ 1 μm resolution) and a selectable wavelength range between 1100 to 200 nm at each pixel. Using synthesized films and heterostructures of transition metal dichalcogenides ($MoS_2$, $WS_2$, $V_xW_{1-x}S_2$, and $WSe_2$), we show that this cost-effective technique detects spectral fingerprints of material identity, doping, grain boundaries, and alloy composition, and enables advanced analysis, including unsupervised machine learning for spatial segmentation.

**Keywords:** Hyperspectral microscopy, differential reflectance, high-throughput characterization, two-dimensional materials, transition-metal dichalcogenides, heterostructures.

## 1. Introduction

Two-dimensional (2D) semiconductors such as the transition metal dichalcogenides (TMDCs, e.g., $MoS_2$, $WS_2$, $MoSe_2$, or $WSe_2$) exhibit appealing properties including good charge transport in sub-nanometer thin films and strong excitonic effects in their optical response, enabling applications in electronic and photonic devices.[1–4] Additionally, lateral and vertical heterostructures result in an even greater variety of interesting properties based on carrier confinement, interlayer charge transfer, moiré superlattices, and other novel effects.[5–7] Early experiments used micrometer-scale TMDC flakes produced by micromechanical exfoliation, limiting device area and scalability. This has motivated significant efforts to synthesize TMDCs deterministically with high-quality over large areas, and to integrate them into heterostructures. However, despite significant progress in the growth of monolayer and few-layer films of TMDCs by techniques like atomic layer deposition (ALD),[8,9] solid-source chemical vapor deposition (SS-CVD),[10–12] metal-organic chemical vapor deposition (MOCVD),[13,14] molecular beam epitaxy (MBE),[15–17] or metal film chalcogenization,[18,19] these techniques do not presently meet the needs of advanced technology nodes, and in many cases have higher defect density than the best exfoliated monolayer films. As a result, the scalable synthesis of high-quality films remains a bottleneck, and many researchers continue to rely upon micromechanical exfoliation.

Optimizing the electrical, optical, and mechanical properties of thin-film TMDCs and other 2D semiconductors requires accurate feedback on sample quality, defined by minimal defects, uniform composition and crystal structure, and precisely controlled doping. Although fabricating and measuring transistors can quantify some key aspects of film quality, the process is time consuming and electrical measurements do not uniquely identify the concentration, location, or type of defects. Additionally, doping or degradation may occur during fabrication,[20,21] which complicates the interpretation of growth quality based on device measurements. Altogether, the slow rate of device fabrication and interpretation is not well-matched to provide timely feedback on high throughput of SS-CVD or MOCVD, making it impractical to fabricate and test devices from every growth batch. Therefore, rapid assessment of crystalline, interfacial, and electrical quality is essential for guiding synthesis optimization and screening samples for further fabrication or advanced measurements.

Many techniques characterize TMDCs after synthesis, but no rapid, widely accepted method exists for overall sample quality assessment. Techniques like scanning or transmission electron microscopy,[22] scanning probe microscopy,[23–25] and confocal optical spectroscopy[26] provide valuable insights but are limited by small fields of view, low throughput, or insufficient sensitivity to the electronic or chemical states. Additionally, individual confocal photoluminescence (PL) or Raman spectra typically measure a small (~ 1 $\mu m^2$) area, and therefore acquisition over many points is needed to determine which data are representative and to identify local fluctuations. Chemically sensitive techniques like x-ray photoelectron spectroscopy (XPS) provide improved composition information but suffer from poor spatial resolution and require high-vacuum conditions, leading to longer measurement times. This underscores the need for a swift, spatially-precise, and non-invasive technique to assess material quality and identify key properties. Optical reflectance, which is sensitive to the electronic structure of TMDCs, can be rapidly measured using hyperspectral optical microscopy to sequentially acquire wavelength-resolved 2D images,[27,28] yet its potential for high-throughput 2D semiconductor characterization remains underexplored.

Here, we demonstrate an optimized hyperspectral optical microscopy technique for rapid characterization of 2D materials and heterostructures (2D-HOM), offering good spatial resolution (< 1 µm) and sensitivity to doping and defects over large areas (~ $300 \times 300\ \mu m^2$) with acquisition times ranging from tens of seconds to several minutes. This system uses wide-field, spectrally-resolved reflectance imaging in a custom optical microscope, spanning wavelengths from 1100 to 200 nm (photon energies of 1.15 to 6.2 eV) to capture electronic transitions in TMDCs such as $MoS_2$, $WS_2$, and $WSe_2$. This method can distinguish between different TMDCs on a substrate, detect spatial variations in doping, identify grain boundaries, map alloy composition, and quickly generate multidimensional data sets for further analysis using machine learning.

## 2. Results and discussions

The 2D-HOM system can be assembled using off-the-shelf optical components. **Fig. 1a** illustrates the 2D-HOM setup showing both the front view and the side view. A commercial tunable light source (Sciencetech TLS-55-X300) directs light from a Xe lamp into a ¼ meter monochromator, outputting a single wavelength within the range of 200 to 2000 nm. The monochromator's output provides good spectral resolution, typically 0.5 to 1 nm full width at half-maximum (FWHM) on the output light. The light is guided by a multimode, solarization-resistant silica fiber and launched into the microscope illumination arm using a reflective collimator. The microscope imaging path is configured for epi-reflectance imaging of opaque samples and uses reflective components to remove chromatic aberration.

The 2D-HOM captures a sequence of reflectance images within the spectral range of 1100 to 200 nm, which is determined by the response of our backside-illuminated silicon CMOS camera (corresponding to energies of 1.15 to 6.2 eV). This yields a three-dimensional (3D) data cube R (x, y, λ) as shown on the left of **Fig. 1b.** The differential reflectance (DR), which is sensitive to a material's electronic structure,[29,30] is calculated from the reflectance of the 2D material regions of interest on the sample ($R_{2D}$) and the reflectance of the substrate ($R_{substrate}$), $\frac{\Delta R}{R} = \frac{R_{2D} - R_{Substrate}}{R_{Substrate}}$ illustrated on the right of **Fig. 1b.** High resolution point spectroscopy across a wide spectral range (hyperspectral) have already been performed to capture wide-field images[31] and determine layer number of the TMDs,[32] demonstrating the usefulness of theses spectra.

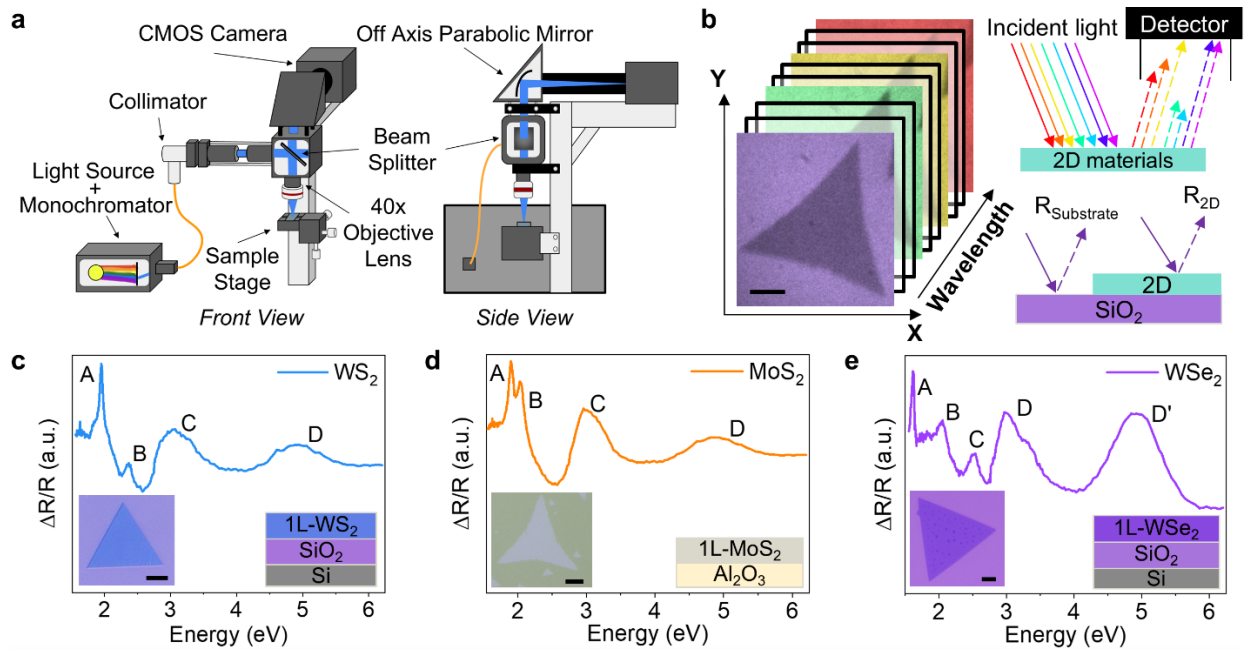

**Figure 1. Design and demonstration of hyperspectral optical microscope for 2D materials (2D-HOM).** **(a)** Illustration of the home-built 2D-HOM system that can measure spectral features from 1.15 eV to 6.2 eV, consisting of three primary parts: (1) the tunable light source with monochromator, providing single wavelength light; (2) the collimator and beam splitter that direct light onto the sample stage through the objective lens; (3) the CMOS camera that collects the reflected light from the sample. **(b)** Schematic of the data cube collected under different wavelengths, where the scale bar is 10 μm. Alongside is the schematic of the reflection that occurs on the surface of the sample. **(c-e)** DR spectra of hybrid-MOCVD[33] (Hy-MOCVD) grown $WS_2$, $MoS_2$, and solid-source CVD grown $WSe_2$. Spectra are collected by averaging signals in the middle region of the materials. The left insets are optical microscope images with scale bars of 5 μm. The right insets are schematic cross-sectional diagrams of the measured sample.

The DR spectra acquired from the 2D-HOM are highly sensitive to the electronic structure of 2D semiconductors. Representative spectra from synthetic CVD monolayers of $WS_2$ (**Fig. 1c**), $MoS_2$ (**Fig. 1d**), and $WSe_2$ (**Fig. 1e**) exhibit excitonic peak features, characteristic of their specific electronic band structure.[34,35] The two lowest energy peaks are the A and B excitons, originating from spin-orbit splitting of the valence band in monolayer TMDCs.[36] The A exciton peak corresponds to the direct band gap transitions at the K point in the Brillouin zone.[34–36] (This is the optical band gap, which is smaller than the electronic gap by the energy of the A exciton.) This peak is typically prominent and plays a major role in our 2D-HOM analysis, although in principle any of the exciton peaks can provide information on the electronic structure. The B exciton peak originates from a similar K-point direct transition from the lower energy level of the spin-orbit split valence band and occurs at slightly higher energies (approximately + 400 meV, + 150 meV and + 440 meV for $WS_2$, $MoS_2$, and $WSe_2$ respectively). The higher-energy exciton peaks are generally associated with optical transitions involving band nesting (C exciton)[37–39] and higher energy transitions near the Γ point (D exciton).[40–42] In $WSe_2$, the spin-split valence band yields an additional high-energy D' exciton.[43] These features provide more nuanced information about the band structure, but further study is required to assess their value for material quality benchmarking. The wide energy range of our system (1.15 to 6.2 eV) enables us to probe many material systems, including multilayers and heterostructures.[44]

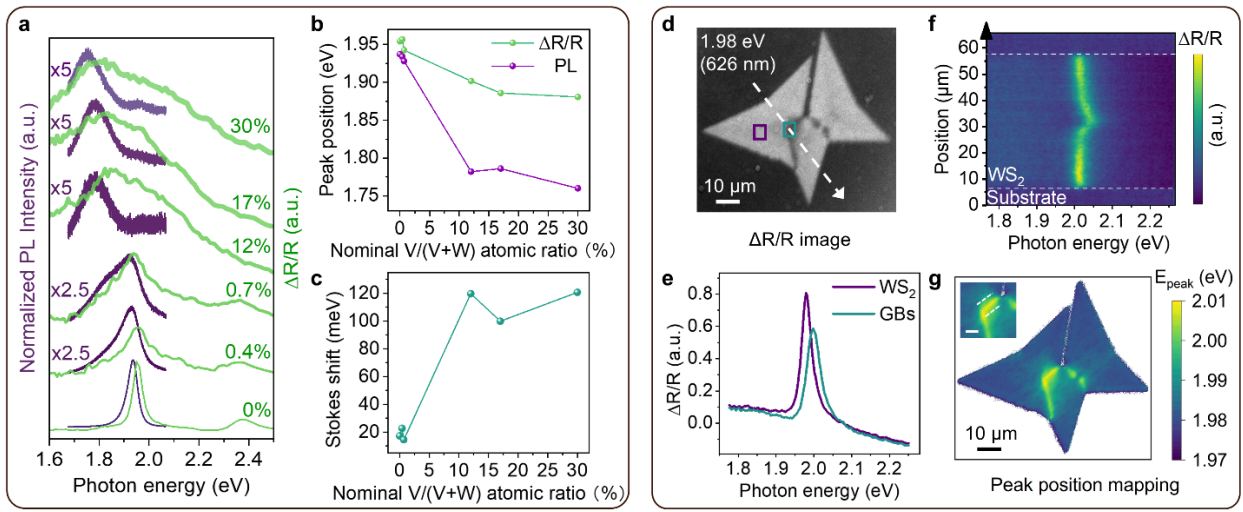

**Figure 2. Analysis of doping and defects using 2D-HOM. (a)** Differential reflectance (DR) and photoluminescence (PL) spectra, in green and purple respectively, of monolayer pure $WS_2$ and V-doped $WS_2$ ($V_xW_{1-x}S_2$) with different nominal doping concentrations (x) from 0.4 % to 30%. For better comparison, the intensity of the PL spectra for x = 12%, 17% and 30% are multiplied by 5, and spectra for x = 0.4% and 0.7% are multiplied by 2.5. **(b)** The PL and DR peak position changes with respect to x. (c) Stokes shift

with respect to x. **(d)** DR image (each pixel has a value of calculated $\frac{\Delta R}{R}$) of a WS$_2$ island with grain boundaries (GBs) under 1.98 eV (626 nm) illumination. **(e)** DR spectra of the selected WS$_2$ region (purple square) and GBs (darker green square). **(f)** Line profile along the slice direction shown as white dotted arrow in **(d)**. **(g)** 2D mapping of the peak positions (fitted using Lorentz function) of the spectra; inset is the magnified GB region with a scale bar of 5 μm.

Electronic structure is a function of composition and is modified by local defects and strain, and 2D-HOM is capable of distinguishing variations and inhomogeneities in the electronic structure which arise from these factors. This can be applied to study impactful challenges in the processing and optimization of 2D semiconductors. For example, substitutional doping requires controlled changes in composition, such as by introducing vanadium as an acceptor in WS$_2$.[45,46] Vanadium-doped WS$_2$ samples (V$_x$W$_{1-x}$S$_2$) with nominal doping concentration x ranging from 0.4% to 30% V were prepared using hybrid MOCVD.[33] As shown by the green curves in **Fig. 2a**, the DR spectra exhibit progressively broadened and redshifted A exciton peaks with increasing vanadium concentration, while the B exciton peaks diminish in prominence. Similarly, the PL spectra (purple in **Fig. 2a**) also broaden and redshift as vanadium content rises. These spectral changes arise from changes to the WS$_2$ electronic structure due to vanadium doping, which increases hole concentration and the density of acceptor states near the valence-band edge.[47] Consequently, both DR and PL spectra experience red shifting and broadening due to narrower bandgap transition energies from excitations involving acceptor states, and PL intensity is quenched by the rise in free holes.[48,49]

Additionally, we correlated doping concentration with defect-related Stokes shifts.[50] As depicted in **Fig. 2b**, the peak positions of both DR and PL decrease with higher nominal vanadium ratios. The Stokes shift (E$_{DR}$ - E$_{PL}$) shown in **Fig. 2c** increases with x, reaching up to 121 meV at 30% V. This trend is consistent with literature where the Stokes shift increases from ~ 40 meV for pure

WS$_2$ to 55 meV for 8 % V doped WS$_2$, indicating that larger Stokes shifts arise from significant doping, suggesting that emission is dominated by additional trapping (acceptor) states.[51,52] The 2D-HOM can detect dopant levels as low as 0.4%, demonstrating sufficient sensitivity for applications such as materials metrology to improve material synthesis process.

We also quantitatively visualized the effects of grain boundaries (GBs), which create strain and one-dimensional lattice defects that modify the local electronic structure.[53,54] As shown in **Fig. 2d**, the DR image of WS$_2$ at 1.98 eV (626 nm) reveals distinct optical contrast between the WS$_2$ regions and the GBs. Such differentiation is challenging to achieve with standard optical microscopy but is important in assessing the quality of a film, especially when polycrystalline edges and facets are not visible. Spectra in **Fig. 2e** are averaged from the highlighted areas in **Fig. 2d** and demonstrate a blue shift and a decrease in amplitude of the DR signal for the A exciton near the GBs. The blue shift is consistent with broadening of the bandgap, which can be attributed to the strain that occurs at the GBs regions due to lattice mismatch.[55] Theoretical models suggest that the conduction band minimum is raised in misoriented GBs under less than 3% compressive strain, increasing the bandgap.[56] This shift of ~ 50 meV is consistent with a ~ 0.6% local compressive strain at the GBs.[57–59] Additionally, 2D-HOM provides detailed spatial characterization of defective regions; the line profile in **Fig. 2f** shows a gradual peak shift from 2 eV to approximately 2.05 eV. By fitting Lorentzian peaks for individual pixels, we obtain a 2D map of A exciton energy in **Fig. 2g**, which demonstrates that GB effects extend around 5 μm beyond the boundary. Evidently, 2D-HOM is capable of assessing the quality of synthetic materials by detecting doping or defects, and may be useful to track perturbations introduced by processing and fabrication.[60]

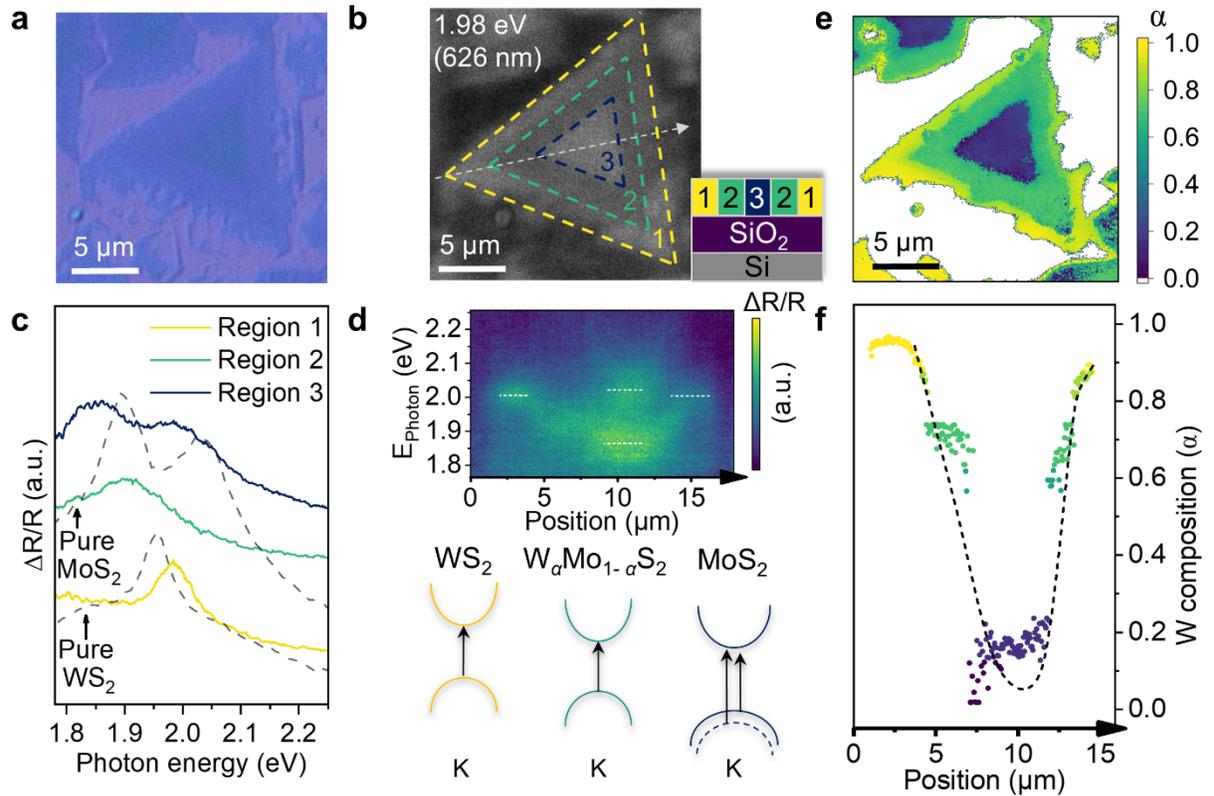

**Figure 3. Analysis of chemical composition using 2D-HOM. (a)** The optical microscope image of a $W_\alpha Mo_{1-\alpha}S_2$ alloy sample transferred onto a SiO$_2$/Si substrate. **(b)** The differential reflectance (DR) image of the alloy sample at 1.98 eV (626 nm). Alongside is the profile structure of the sample. **(c)** DR spectra from regions 1, 2 and 3 labeled in **(b)**. The dotted line represents DR spectra of pure WS$_2$ (bottom) and MoS$_2$ (top) respectively. **(d)** The upper part is the line profile of the slice direction shown as white dotted arrow in **(b)**. The lower part illustrates the possible band structure changes in the alloy sample. **(e)** 2D mapping of W composition (α) of the same triangle in **(a)**. **(f)** Calculated W composition (α) line profile along slice direction marked by the arrow in **(b)**. The dashed line is to guide the eye.

Local variations in composition can be inferred through 2D-HOM analysis. To illustrate this, we analyze an alloyed lateral heterostructure sample with a composition of $W_\alpha Mo_{1-\alpha}S_2$. The sample's profile is depicted in the lower right of **Fig. 3b**, with regions 1, 2, and 3 labeled on the right. Under optical microscope (**Fig. 3a**), the alloy appears as a continuous triangle with minimal internal contrast, although a subtle central triangular region is discernible. The DR image (**Fig. 3b**) at 1.98 eV (626 nm) clearly visualizes three distinct regions – outer (R1), middle (R2), and inner (R3) – with varying optical contrast.

DR spectra for each region are shown in **Fig. 3c** alongside pure $WS_2$ and $MoS_2$ spectra (dashed lines). R1 exhibits an A exciton peak at ~ 2 eV which resembles pure $WS_2$, and R3 displays A and B exciton peaks at 1.86 eV and 2 eV, respectively, which resemble pure $MoS_2$.[35] The middle, R2, exhibits a peak around 1.9 eV, which is consistent with a $W_\alpha Mo_{1-\alpha}S_2$ alloy. A DR line profile, shown in **Fig. 3d**, was extracted from the white dotted line in **Fig. 3b**, illustrating the gradual shift in the A exciton position from R1 to R2, and the emergence of a second peak, consistent with the $MoS_2$-like B exciton, in the transition from R2 to R3.[61,62]

Using this electronic structure information, we inferred the W concentration (α in $W_\alpha Mo_{1-\alpha}S_2$) along the line profile (**Fig. 3f**) and in a 2D map (**Fig. 3e**) by interpolating between the measured $WS_2$ and $MoS_2$ A and B exciton peak positions, while considering the band bowing effect, $E_{W_\alpha Mo_{1-\alpha}S_2} = \alpha E_{WS_2} + (1-\alpha) E_{MoS_2} - b * \alpha(1-\alpha)$. The *b* in the equation is the bowing factor, 0.25 eV and 0.19 eV for A and B excitons of TMDs respectively.[63] Along the slice direction (white dashed line) in **Fig. 3b**, the W composition (α) decreases from α = 0.95 near the corner to ~ 0.1 at the center of the triangle, then rises again to ~ 0.9 near the edge. The 2D mapping further highlights this compositional variation from R1 (yellow) to R3 (dark blue), reflecting a decrease in W composition (α). These results align with our previous finding that growth of alloys with a Mo:W precursor ratio of 1:2 typically exhibits a $MoS_2$-like core and a $WS_2$-like shell.[33] 2D-HOM adds the capability for quick and quantitative spatial composition analysis, which is a useful addition to existing characterization techniques for 2D heterostructure growth. This alloy sample demonstrates that 2D-HOM can quantitatively distinguish chemical compositions in a lateral heterostructure using the extracted spectral information. Similarly high spatial resolution DR mapping is expected to be useful for large-area analysis of materials quality, inhomogeneity, and in-situ composition changes during synthesis.

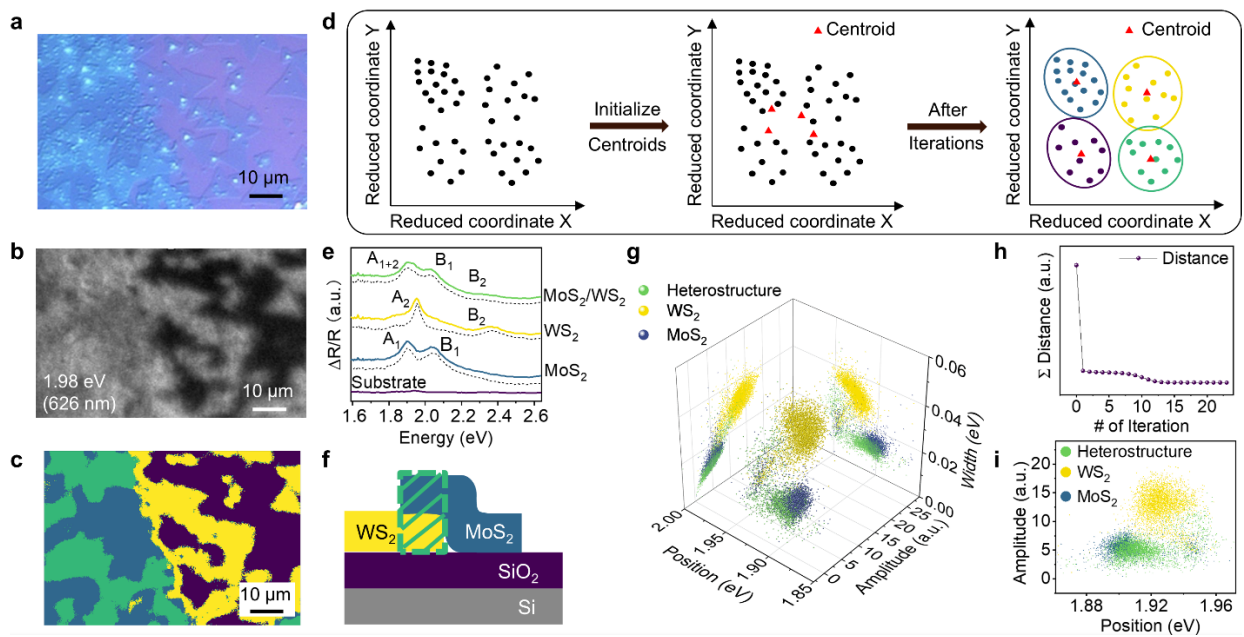

**Figure 4. Potentials of 2D-HOM in artificial intelligence. (a)** Optical microscope image of $WS_2/MoS_2$ vertical heterostructure with a schematic profile shown in **(f)**. The scale bar is 10 μm. **(b)** Differential reflectance image of the heterostructure at 1.98 eV. **(c)** Image formed by segmenting the heterostructure region via K-means clustering, using the 3D DR data cube as the input. **(d)** Illustration of K-means clustering algorithm. The red triangles demonstrate different centroids. **(e)** Plot of the characteristic exemplar spectra extracted for each of the regions defined by K-means clustering. Corresponding dotted lines are the manually selected spectra of materials region. **(g)** 3D scatterplot of parameters for peaks fitted to the A-exciton features of the individual spectra at each (x, y) coordinate. The data points are labeled with colors corresponding to their K-means cluster assignment. This 3D display indicates that clustering of the raw hyperspectral data accurately differentiates the individual sample regions. The left and right walls of the 3D plot area show projections of the scatterplot for the Amplitude-Width plane and the Position-Width plane, respectively. **(h)** Plot of changes in the summation distance from all points to the centroid as a function of the number of K-means clustering iterations, demonstrating convergence during the clustering process. **(I)** Plot of the fitted A-exciton peak parameters projected to the Position-Amplitude plane.

Beyond the manual spectral analysis outlined above, the 2D-HOM system provides a rich combination of spatial and spectral data that can be mined for further information using emerging artificial intelligence (AI) or machine learning (ML) techniques. As an illustrative example where the categorization capabilities of AI/ML algorithms prove useful, we studied a $WS_2/MoS_2$ vertical heterostructure. As shown in **Fig. 4a** and **Fig. 4f**, we fabricated the $WS_2/MoS_2$ vertical heterostructure by transferring Hy-MOCVD-grown thin films (see methods section) and acquired a hyperspectral DR dataset (**Fig. 4b**). We then applied a K-Means clustering algorithm (**Fig. 4d** and method section) directly to the heterostructure's 3D DR data cube, partitioning each pixel into one of the clusters based on its spectral characteristics according to their squared Euclidean distance to the cluster centroids. The centroids are updated iteratively until the cluster assignment converges.

We assigned four clusters to this system based on our prior knowledge of the input materials: $WS_2$, $MoS_2$, heterostructure regions, and bare substrate. The clustering results, shown in **Fig. 4c**, categorize regions into four distinct clusters: purple ($SiO_2$ substrate), yellow (pure $WS_2$), dark blue (pure $MoS_2$), and green ($WS_2/MoS_2$ vertical heterostructure). Exemplar spectra of these four regions are shown in **Fig. 4e**. The substrate spectrum is flat, while the peak positions of the $WS_2$ and $MoS_2$ spectra match expected values.[35,36] The heterostructure spectrum features a broadened A exciton peak, representing a combination of $MoS_2$ ($A_1$) and $WS_2$ ($A_2$) excitons, and dual B exciton peaks corresponding to $MoS_2$ ($B_1$) and $WS_2$ ($B_2$). The asymmetry in the $B_2$ peak arises from the presence of the $A_2$ peak. This aligns with previous studies indicating that the total DR response of layered heterostructures can be modeled as a linear combination of their constituent layers.[29] Additionally, manually selected spectra for each region (excluding the substrate) are overlaid as black dotted lines in **Fig. 4e**, further validating the clustering accuracy.

To further substantiate our clustering approach, **Fig. 4g** presents a 3D visualization of the clustering outcomes. For each pixel within the materials clusters, the A exciton peak (lowest energy) was fitted using a Lorentz function, extracting three parameters: peak position, amplitude and width, plotted in **Fig. 4g**. The separation of the data points in the 3D view corroborates the effectiveness of the clustering. **Fig. 4h** illustrates the total cumulative Euclidean distance between each data point and its corresponding centroid versus the number of iterations, demonstrating that the algorithm successfully converges by minimizing these distances.

The robustness of the clustering is evident in the peak position-amplitude plane in **Fig. 4I**: although the overlapping peak positions of $MoS_2$ and the heterostructure would complicate traditional segmentation, the added dimensionality from clustering enables clear separation of these regions. This demonstrates the potential of 2D-HOM to leverage AI/ML for high-throughput characterization, enhancing the analysis of complex hyperspectral datasets and revealing subtle features that may escape manual interpretation.[64]

## 3. Conclusion

In this study, we demonstrate that 2D-HOM enables rapid and versatile characterization of 2D materials across a broad spectral range, from the deep ultraviolet to near-infrared. By capturing rich 3D hyperspectral data sets, 2D-HOM enables identification of materials based on their spectral signatures and reveals spatial variations in composition or defect density. These capabilities make 2D-HOM a powerful platform for fast quality assessment and inhomogeneity detection. When integrated with AI/ML algorithms, 2D-HOM can accurately differentiate regions within complex heterostructures – a task that is challenging for conventional optical microscopes. Additional dimensions of contrast could be introduced through electrostatic gating,[65,66] photoexcitation,[67] or wavelength-selective detection[68]. These results highlight the potential of 2D-HOM for high-throughput materials evaluation and support future advances in characterization and synthesis of 2D materials.[69]

## 4. Methods

**Materials growth**

Monolayer $WS_2$, $MoS_2$, $W_\alpha Mo_{1-\alpha}S_2$ alloy, and V-doped monolayer $WS_2$ ($V_xW_{1-x}S_2$) were prepared with dip-coating Hy-MOCVD, as reported in our previous paper.[33] For the growth of monolayer $WS_2$ on the $SiO_2$/Si substrate, ammonium metatungstate hydrate (AMT) and potassium hydroxide (KOH) solution (0.6 g + 0.1 g in 30 mL deionized water) was used as tungsten source and 0.12 sccm diethyl sulfide (DES) was used as sulfur source. For the growth of monolayer $MoS_2$ on c-plane sapphire substrate, ammonium molybdate (AMM) and KOH solution (0.43 g + 0.1 g in 30 mL deionized water) was used molybdenum source and 0.05 sccm DES was used as sulfur source. In the growth of $W_\alpha Mo_{1-\alpha}S_2$ alloy on c-plane sapphire substrate, the solution of AMT, AMM and KOH with Mo/W mole ratio of 1/2 was used as metal source and 0.05 sccm DES was as sulfur source. The alloy sample was transferred from sapphire to a $SiO_2$/Si substrate using a PMMA-assisted wet transfer method for 2D-HOM imaging. For V-doped monolayer $WS_2$ on $SiO_2$/Si, ammonium vanadate, AMT and KOH solution was used as metal source and 1.2 sccm DES was used as sulfur source. The mole ratios of V/(V+W) in the metal solution were used to determine the nominal doping concentrations of V in the samples. Other growth parameters for the samples were identical. 1600 sccm Ar and 1 sccm $H_2$ were used as carrier gases. These Hy-MOCVD growths were completed in a 125 mm inner diameter tube furnace at 775 °C for 6 hours.

The $WSe_2$ crystals were grown using a two-zone solid source CVD method. Before the growth process, 2 g of selenium (purity 99.999%, Thermo Fisher) was placed at the center of the first zone, while 8 mg of tungsten (VI) oxide (purity 99.995%, Sigma Aldrich) was placed in the second zone, which was positioned 30 cm away from the selenium boat. A 96-nm thick $SiO_2$ on Si substrate was then placed 5 cm downstream from the $WO_3$. The system was purged down to a base pressure of 10 mTorr before being raised to atmospheric pressure using Ar, and this pressure was maintained throughout the process. The temperature of the $WO_3$ and Se zones was set at 880 °C and 500 °C, respectively. The growth process was carried out for 40 minutes using a mixture of 25 sccm of Ar and 5 sccm of $H_2$. After the growth, the furnace was naturally cooled down to room temperature.

**Preparation of $WS_2/MoS_2$ heterostructure**

The $WS_2/MoS_2$ heterostructure was prepared by stacking $MoS_2$ film onto as-grown $WS_2$ triangles on a $SiO_2/Si$ substrate, using a wet transfer method. The $MoS_2$ film separately grown on a sapphire substrate was spin-coated with PMMA, and then slowly dipped into deionized water. The $MoS_2$/PMMA layer was delaminated from sapphire by the deionized water and then transferred onto $WS_2$ as-grown on $SiO_2/Si$. The PMMA was removed by immersing the sample in acetone for 15 mins.

**K-means cluster algorithm**

The K-Means clustering algorithm, implemented in the scikit-learn package[70], is designed to segregate three-dimensional data into n specified clusters ($C_1$, $C_2$, . . ., $C_n$), with each cluster $C_n$ containing a set of selected spectra with the number of $m_n$. In the 3D data generated by 2D-HOM, there are x*y total spectra, each providing the reflectance information within the whole wavelength ranges, which have P variable-spectral points. Here, x and y represent the number of pixels of the 2D plane. The centroid of a cluster $C_k$ is a point residing in a P-dimensional space, and it is computed by averaging the values of each variable across the objects within that cluster. For instance, the centroid value for the $j^{th}$ variable in cluster $C_n$ is calculated as follows:

$$\bar{x}_j^{(n)} = \frac{1}{m_n} \sum_{i \in C_n} x_{ij}$$

And the centroid cluster of $C_n$ is:

$$\bar{x}^{(n)} = (\overline{x_1}^{(n)}, \overline{x_2}^{(n)}, \ldots \overline{x_p}^{(n)})$$

The standard k-means clustering algorithm functions through the following iterative procedure:

(1) The squared Euclidean distance between the $i^{th}$ object and the nth seed vector is obtained by

$$d^2(i,n) = \sum_{j=1}^{p}(x_{ij} - s_j^{(n)})^2$$

(2) Following the initial allocation of hyperspectral spectra to clusters, compute the cluster centroid for each cluster. All the spectra are then compared to these centroids using the squared Euclidean distance $d^2(i, n)$. Each spectrum is moved to the cluster whose centroid is the closest, effectively reassigning spectra to the cluster with the most similar centroid.

(3) Using the updated cluster ID resulting from the previous step, determining the new centroid for each cluster by recalculating the centroids after all spectra have been reassigned to their respective clusters.

(4) Steps 2 and 3 are iteratively executed until no further changes occur in the composition of each cluster. In other words, this process is repeated until it converges, and each spectrum remains in its assigned cluster without further reassignment.

**Materials characterization and analysis**

Optical microscope images were taken using an Olympus BX-51 microscope in epi-reflection geometry. Photoluminescence (PL) spectra were collected at room temperature using 532 nm laser excitation with a HORIBA Scientific LabRAM HR Evolution confocal microscope. Differential reflectance (DR) images are generated by first selecting a nearby substrate region and averaging signal within it as $R_{Substrate}$ and then applying this value throughout the entire image to calculate the DR at each pixel. The DR curves are fitted with a single Lorentzian peak after subtracting a linear background. The PL curves are fitted by first subtracting a linear background and then performing a fit using a weighted Gaussian-Lorentzian function consisting of a blend of 55%

Gaussian and 45% Lorentzian character. PL curves of 0%, 0.4%, 0.7% V-doped samples are fitted using 2 peaks and the higher energy peaks are chosen to calculate the Stokes shift. PL curves of 12%, 17%, 30% V-doped samples are fitted using 1 peak.

**Data and code availability**

The code and data that support the findings of this study will be made freely available upon publication through GitHub and Zenodo (DOI:10.5281/zenodo.15128160), respectively.

**5. Acknowledgments**

Primary support for this work was provided by the National Science Foundation Future of Semiconductors program (FuSe2, Award 2425218; Z.P., A.U., A.J.M., E.P.) and the Office of Naval Research Young Investigator Program (Award N000142512137; X.Z.). Sample growth received additional funding from the Stanford SystemX Alliance, sponsored by Taiwan Semiconductor Manufacturing Company (A.–T.H.). Early development was supported by the U.S. Department of Energy, Office of Basic Energy Sciences, under Contract DE-AC02-76SF00515 (Z.P., A.J.M.), and by instrumentation acquired under NSF MRI Award 2018008. Portions of this work were performed at the Stanford Nano Shared Facilities (SNSF, RRID: SCR_023230), supported by NSF Award ECCS-2026822.